\documentstyle[12pt]{article}
\input{epsf}
\newcommand{\beq}{\begin{equation}}
\newcommand{\eeq}{\end{equation}}
\newcommand{\be}{\begin{eqnarray}}
\newcommand{\ee}{\end{eqnarray}}
\begin{document}
\rightline{RUB-TPII-8/98}
\rightline{hep-ph/9807229}
\vspace{1cm}
\begin{center}
{\bf\LARGE Pion and photon light--cone wave functions from the instanton 
vacuum}
\\[1.5cm]
{\bf V.Yu.\ Petrov$^{\rm 1}$, M.V.\ Polyakov$^{\rm 1, 2}$,
R.\ Ruskov$^{\rm 3}$, C.\ Weiss$^{\rm 2}$ and K. Goeke$^{\rm 2}$}
\\[0.2cm]
$^{\rm 1}${\em Theory Division of Petersburg Nuclear Physics Institute \\
188350 Gatchina, Leningrad District, Russian Federation}
\\[0.3cm]
$^{\rm 2}${\em Institut f\"ur Theoretische Physik II \\
Ruhr--Universit\"at Bochum \\
D--44780 Bochum, Germany}
\\[0.3cm]
$^{\rm 3}${\em Bogoliubov Laboratory of Theoretical Physics \\
Joint Institute for Nuclear Research \\
141980 Dubna, Russian Federation}
\end{center}
\vspace{1cm}
\begin{abstract}
\noindent
The leading--twist wave functions of the pion and the photon at a low
normalization point are calculated in the effective low--energy theory
derived from the instanton vacuum. The pion wave function is found to
be close to the asymptotic one, consistent with the recent CLEO
measurements. The photon wave function is non-zero at the endpoints.
This different behavior
is a consequence of the momentum dependence of the dynamical quark 
mass suggested by the instanton vacuum. We comment on the relation
of meson wave functions and off-forward parton distributions
in this model.
\end{abstract}
\vspace{1.5cm} PACS: 12.38.Lg, 13.60.Fz, 13.60.Le
\\
Keywords: \parbox[t]{13cm}{light--cone wave functions,
exclusive reactions, non-perturbative methods in QCD,
chiral symmetry, instanton vacuum}
\newpage
\noindent
Hadron light--cone wave functions (also called distribution amplitudes)
parametrize the non-perturbative information entering in the amplitudes
for exclusive hard scattering processes in QCD
\cite{BL79,EfrRad80,ChZh77}.
The pion wave function enters in the description of the pion electromagnetic
form factor and pion--meson transition form factors (for a review 
see \cite{ChZh84}), and in exclusive pion production in photon--photon
\cite{Ong95,JKR96,KrollRaulfs96,RR96,MusRad97} and
photon--nucleon processes \cite{MPW97}. The photon wave function 
appears, for instance, in radiative hyperon decays \cite{BBK89}, or 
in photon--photon processes where the virtuality of one of the photons is 
small, such as $\gamma\gamma^\ast \rightarrow \pi^0$ \cite{RR96} or 
power--suppressed contributions to 
deeply--virtual Compton scattering \cite{R96,Ji97}.
\par
A calculation of the meson and photon wave functions from first
principles requires a theory of the non-perturbative effects giving rise
to hadron structure. Meson wave functions have extensively been studied
using QCD sum rules. The original suggestion
by Chernyak and Zhitnitsky of a ``double--humped'' wave function of the pion
at a low scale, far from the asymptotic form, was based on an extraction of 
the first few moments from a standard QCD sum rule 
approach \cite{ChZh84},
which has been criticized and revised in Refs.\cite{MR86,BF89}.
Additional arguments in favour of a form of the
pion wave functions close to the asymptotic one
came from the analysis of the transition form 
factor $\gamma\gamma^\ast \rightarrow \pi^0$ \cite{RR96}. The recent
measurements of this form factor by the CLEO collaboration are
consistent with a near--asymptotic form of the wave 
function \cite{CLEO98}.
\par
About the leading--twist wave function of
the soft photon little is known either from phenomenology or from first
principles. It has been discussed {\em e.g.}\ in an analysis of
photon--meson transition form factors in the framework of a constituent
quark model by Anisovich {\em et al.}  \cite{AMN96}.
\par
In this paper we study the pion and photon wave functions in the
instanton vacuum. The picture of the QCD vacuum as a dilute medium of
instantons explains the dynamical breaking of chiral symmetry, which is
the non-perturbative phenomenon most important for hadron structure at
low energies \cite{DP86}. Quarks interact with the fermionic zero modes
of the individual instantons in the medium, which leads to the formation
of a chiral condensate. One derives from the instanton vacuum an
effective theory of quarks with a dynamical mass which drops to zero at
Euclidean momenta of the order of the inverse average instanton size,
$\bar\rho^{-1} \simeq 600 \, {\rm MeV}$ \cite{D96}.  
\par 
The pion and photon 
wave functions can be extracted from correlation
functions of light--ray operators with the mesonic {\em viz.}\
electromagnetic current, which can be computed in the effective low--energy
theory.  The normalization point of the wave functions obtained in this
approach is of the order of $\bar\rho^{-1} \simeq 600 \, {\rm MeV}$.
The pion wave function has been computed in Ref.\cite{PP97}; 
it was found to be close to the asymptotic
one. Our intention here is twofold. First, we wish to expand the
investigation of the pion wave function, discuss its scale 
dependence and compare with the recent CLEO measurements \cite{CLEO98}.
Second, we compute also the photon wave function. In particular, we
shall be interested in comparing the photon and the pion wave
functions. As will be seen below, the two exhibit different behavior at
the endpoints, $u \rightarrow 0$ and $1$.  This is a consequence of the
momentum--dependence of the dynamical quark mass implied by the instanton
vacuum.
\par
Additional motivation for studying the photon and pion wave
function comes from their importance for deeply--virtual Compton
scattering and hard meson production \cite{MPW97,R96,Ji97}.  The 
factorization of the Compton amplitude in the deeply virtual domain 
involves the off-forward parton distributions (OFPD'S) of the
nucleon. Recently, Radyushkin has argued, on the basis of a ``meson
exchange'' contribution to the so--called double distributions related to 
the OFPD's \cite{Radyushkin98}, that the OFPD's may be discontinuous at 
$x = \pm \xi / 2$ ($\xi$ is the longitudinal component of the momentum
transfer to the nucleon) if the meson wave function were non-zero at the
end points.  Such discontinuities had indeed been observed in a
calculation of the isosinglet OFPD's of the nucleon in the effective
low--energy theory derived from the instanton vacuum, where the nucleon
is described as a chiral soliton \cite{PPPBGW97}. It was seen there that
near $x = \pm \xi / 2$ the behavior of the OFPD is governed by the
momentum dependence of the dynamical quark mass, which turns the
would--be discontinuity into a sharp but continuous crossover. Here we
shall see that the same physical mechanism --- the momentum dependence
of the dynamical quark mass obtained from the instanton vacuum --- is
responsible for the endpoint behavior of the meson wave function,
showing that this approach provides a consistent realization of 
Radyushkin's general arguments.
\par
{\em Pion and photon wave function.}
The basic objects arising in the factorization of hard scattering
amplitudes involving mesons or photons in the initial or final state are
matrix elements of certain gauge--invariant non-local operators between
the meson (photon) states and the vacuum. Their classification in
structures of different twist and their respective role in the
asymptotic limit has been discussed {\em e.g.}\ in
Ref.\cite{ChZh84}.
The twist--2 wave function of the pion is defined through the matrix
element
\be
\langle 0 | \bar d (z) \gamma_\mu \gamma_5 [z, -z] u (-z)
| \pi^+ (P ) \rangle
&=& i \sqrt{2} F_\pi P_\mu
\int_0^1 du \; e^{i (2 u - 1) P\cdot z} \phi_\pi (u) .
\label{phi_pion}
\ee
Here $z$ is a light--like 4--vector ($z^2 = 0$), and
\be
[z, -z] &\equiv& \mbox{P}\, \exp \left[
\int_{-1}^1 dt\; z^\mu A_\mu (t z)
\right]
\label{P_exp}
\ee
denotes the path--ordered exponential of the gauge field, required by
gauge invariance; the path here is defined to be along the light--like
direction $z$. Furthermore, $P$ is the pion 4--momentum; we shall consider 
the chiral limit, $P^2 = 0$. 
Finally, in Eq.(\ref{phi_pion}) $F_\pi$ denotes the
usual weak pion decay constant,
\be
\langle 0 | \bar d (0) \gamma_\mu \gamma_5 u (0)
| \pi^+ (P ) \rangle &=& i \sqrt{2} F_\pi P_\mu 
\label{fpi_def}
\ee
($F_\pi = 93\, {\rm MeV}$), and the wave function is 
normalized according to
\be
\int_0^1 du \; \phi_\pi (u) &=& 1 .
\label{normalization}
\ee
\par
The matrix element in Eq.(\ref{phi_pion})
involves a chiral--even operator with Dirac matrix
$\gamma_\mu \gamma_5$; the corresponding matrix element with a chiral--odd 
operator ($\gamma_5$) is of higher twist. 
\par
The twist decomposition of the matrix elements for the photon parallels 
that for the rho meson (see {\em e.g.} \cite{BBKT98}),
the only difference --- except isospin --- being that the real photon has 
only two transverse polarization states.  The twist--2 wave function of the
photon is defined by the matrix element
\be
\langle 0 | \bar u (z) \sigma_{\mu\nu} [z, -z] u (-z)
| \gamma (P, \lambda ) \rangle
&=&
i \, {\textstyle\frac{2}{3}} f_{\gamma\perp}
\left( e^{(\lambda )}_{\perp\mu} P_\nu -
e^{(\lambda )}_{\perp\nu} P_\mu \right)
\int_0^1 du \; e^{i (2 u - 1) P\cdot z} \phi_{\gamma\perp} (u)
\nonumber \\
&& \;\; + \;\; \mbox{higher twists} ,
\label{phi_photon_trans}
\ee
Here the photon state is characterized by the four--momentum,
$P$ ($P^2 = 0$), and polarization vector $e^{(\lambda )}_\perp$,
where $e^{(\lambda )}_\perp$ is transverse with respect to $z$ and $P$
($z\cdot P \neq 0$).
\par
We have not written explicitly in Eq.(\ref{phi_photon_trans})
the terms with tensor structures corresponding to contributions
of twist 3 and 4.
The matrix element of the chiral--odd operator with $\sigma_{\mu\nu}$
represents the only
possible twist--2 matrix element for the on--shell photon; a twist--2,
structure with a chiral--even operator ($\gamma_\mu$) is possible
only for a virtual photon or rho meson with longitudinal
polarization.
\par
In Eq.(\ref{phi_photon_trans}), the normalization constant,
$f_{\gamma\perp}$, is defined through the matrix element of the
corresponding local operator ($z = 0$),
\be
\langle 0 | \bar u (0) \sigma_{\mu\nu} u (0)
| \gamma (P, \lambda ) \rangle
&=& i\, {\textstyle\frac{2}{3}}
f_{\gamma\perp} \left( e_{\perp\mu}^{(\lambda )} P_\nu 
- e_{\perp\nu}^{(\lambda )} P_\mu \right) ,
\label{f_gamma_def}
\ee
so that the photon wave function is normalized analogous to
Eq.(\ref{normalization}).
\par
{\em Effective low--energy theory from the instanton vacuum.}
We now compute the matrix elements Eqs.(\ref{phi_pion}) and
(\ref{phi_photon_trans}) in the effective low--energy theory which has been
derived from the instanton vacuum in the large $N_c$--limit.  This
effective theory describes the interaction of a quark field, $\psi$,
with a pion field, $\pi$, by an effective action
\be
S_{\rm eff} 
&=& \int d^4 x \;
\bar\psi (x) \left[ i\gamma^\mu \partial_\mu - M U^{\gamma_5} (x) \right]
\psi (x) ,
\label{L}
\ee
where
\be
U^{\gamma_5}(x) &=& \exp \left[ i \gamma_5 \tau^a \pi^a (x) \right] .
\ee
Note that the interaction is chirally invariant. Here, $M$ denotes the
dynamical quark mass, which appears due to the spontaneous breaking of
chiral symmetry. On general grounds, this effective theory is valid up
to some ultraviolet cutoff. In the instanton vacuum this cutoff is
implemented in the form of a specific momentum dependence of the
dynamical quark mass, which leads to the following form of the
quark--pion coupling:
\be
\int d^4 x \; \bar\psi (x) \, M U^{\gamma_5} (x) \, \psi (x)
&\rightarrow& M \int\frac{d^4 k}{(2\pi )^4} \int\frac{d^4 l}{(2\pi )^4}
\bar\psi (k) F(k) U^{\gamma_5} (k - l) F(l) \psi (l) .
\label{quark_pion_coupling}
\ee
Here $F(k)$ is a form factor, $F(0) = 1$, related to the Fourier
transform of the instanton zero mode, which drops to zero for space-like
momenta larger than the inverse average instanton 
size\footnote{Here and in the following, all momenta refer to the
Minkowskian metric.},
\be
F(k) &\rightarrow 0& \hspace{1cm} \mbox{for} \hspace{1cm}
-k^2 \gg \bar\rho^{-2} .
\ee
We note that, when working with the effective theory, Eq.(\ref{L}), it is
assumed that the dynamical quark mass is parametrically small compared
to the ultraviolet cutoff; their ratio is proportional to the packing
fraction of instantons in the vacuum,
\be
(M \bar\rho )^2 \propto \left( \frac{\bar\rho}{\bar R} \right)^4 .
\ee
\par
In order to compute the photon wave function we need to couple an
electromagnetic field to the quark fields of the effective Lagrangian,
Eq.(\ref{L}). The interaction of the electromagnetic field with the
quarks is dominated by the pointlike interaction which derives from
the kinetic term in the effective Lagrangian. The non-pointlike coupling
arising from the momentum--dependent mass term, which would involve
derivatives of the form factor $F(k)$, is parametrically suppressed in
$M\bar\rho$ relative to the pointlike one. Thus, in leading order
in $M\bar\rho$ we shall work with the point--like electromagnetic
current ($\hat{Q}$ is the quark charge matrix)
\be
J_\mu^{\rm e.m.} (x) &=& \bar\psi (x) \gamma_\mu \hat{Q} \psi (x) .
\label{J_em}
\ee
One should note the different role of the form factors in the coupling
of the pion and the electromagnetic field to the quarks.  While in the
pion--quark coupling, Eq.(\ref{quark_pion_coupling}), the form factors
act {\em multiplicatively}, suppressing the coupling for large quark
virtualities, in the case of the photon the form factors inherent in
Eq.(\ref{L}) make only an {\em additive} (and parametrically small)
contribution to the point--like coupling.  This has important
consequences for the behavior of the pion and photon wave functions, in
particular at the end points $u \rightarrow 0$ and $1$.
\par
{\em Computation of wave functions in the effective low--energy theory.}
To compute the photon wave function, Eq.(\ref{phi_photon_trans}), we
proceed in analogy to the calculation of the pion wave function in
Ref.\cite{PP97}. The matrix element between the one--photon state and
the vacuum is extracted from the correlation function of the light--ray
operator with the electromagnetic current,
\be
i \int d^4 x \; e^{-i P\cdot x} \langle 0 | \mbox{T}
\left\{ J_\rho^{\rm e.m.} (x) , \bar u (z)
\sigma_{\mu\nu} [z, -z] u(-z) \right\} | 0 \rangle
\label{correlator}
\ee
When computing this correlation function in the instanton vacuum we can
drop the path--ordered exponential in the twist--2 operator, since its
contribution is parametrically of order
$(\bar \rho / \bar R)^4 \propto (M\bar\rho )^2$ (see
Refs.\cite{DPPPW96,PW97} for a discussion).  The
correlator can then be evaluated in the effective low--energy theory
given by Eq.(\ref{L}). In the large--$N_c$ limit, Eq.(\ref{correlator}) 
is given by a simple quark loop, with the quark propagator subject
to the momentum--dependent dynamical quark mass, {\em cf.}\ Eq.(\ref{L}).
Projecting on the twist--2 structure and contracting with
the photon polarization vector, we obtain for the matrix element on the
R.H.S.\ of Eq.(\ref{phi_photon_trans}):
\be
\lefteqn{ z^{\nu}\; 
\langle 0 | \bar u (z) \sigma_{\mu\nu} u (-z) | \gamma (P, \lambda )
\rangle \;\; = \;\;
- {\textstyle\frac{8}{3}} N_c M e_{\perp\mu}^{(\lambda )} } &&
\nonumber \\
&&\times
\int\frac{d^4 k}{(2\pi )^4} e^{-i (P - 2 k) \cdot z} D(k) D(k - P)
\left[z \cdot (P - k) \, F^2 (k) + z \cdot k \, F^2 (k - P) \right] ,
\label{photon_me}
\ee
where we have set
\be
D(k) &=& \frac{1}{k^2 - M^2 F^4 (k) + i0} .
\ee
Note that we take into account here the form factors ({\em i.e.}, the
momentum dependence of the dynamical quark mass) in the denominators of
the quark propagators.
\par
To evaluate the integral in Eq.(\ref{photon_me}) it is convenient to
introduce light--like vector components. Let $n$ and $\tilde{n}$ be
dimensionless light--like vectors parallel to $z$ and $P$, with
\be
n \cdot \tilde{n} &=& 2.
\ee
Then we can decompose
\be
k_\mu &=& \frac{k^+}{2} n_\mu + \frac{k^-}{2} \tilde{n}_\mu + k^\perp_\mu
\hspace{2cm}
k^+ \;\equiv\; n\cdot k , \hspace{.5cm} k^- \; \equiv \;
\tilde{n} \cdot k .
\ee
Expressing the integral in Eq.(\ref{photon_me}) in terms of the
light--like vector components, and inserting Eq.(\ref{photon_me}) in
Eq.(\ref{phi_photon_trans}), we can read off the expression for the
photon wave function:
\be
\phi_{\gamma\perp} (u)
&=& (- i) \frac{2 N_c M P^+}{f_{\gamma\perp}}
\int\frac{d^2 k^\perp}{(2\pi )^2}
\int\frac{dk^-}{2\pi} \int\frac{dk^+}{2\pi} \;
\delta ( k^+ - u P^+ )
D(k) D(k - P)
\nonumber \\
&& \times
\left[ (1 - u) F^2 (k) + u F^2 (k - P) \right] .
\label{phi_photon_integral}
\ee
The normalization constant can be obtained by computing the matrix
element with the local operator, Eq.(\ref{f_gamma_def}).  After taking
the limit $P^2 \rightarrow 0$ one finds
\be
f_{\gamma\perp} &=& (-i) 4 N_c M
\int\frac{d^4 k}{(2\pi )^4} \; D^2 (k) \;
\left[ F^2 (k) - \frac{k^2}{2} \frac{dF^2 (k)}{d(k^2)} \right] .
\label{normalization_integral}
\ee
Integrating Eq.(\ref{phi_photon_integral}) over $u$ and using the fact
that the integral does not depend on the light--cone vector $n$, one
easily verifies that Eq.(\ref{phi_photon_integral}) obeys the
normalization condition, Eq.(\ref{normalization}).
\par
Eq.(\ref{phi_photon_integral}) should be compared to the result for the
pion wave function, Eq.(\ref{phi_pion}), which was derived in
Ref.\cite{PP97}:
\be
\phi_\pi (u)
&=& (-i) \frac{2 N_c M^2 P^+}{F_\pi^2}
\int\frac{d^2 k^\perp}{(2\pi )^2}
\int\frac{dk^-}{2\pi} \int\frac{dk^+}{2\pi} \;
\delta ( k^+ - u P^+ ) D(k) D(k - P)
\nonumber \\
&& \times  F(k) F(k - P)
\left[ (1 - u) F^2 (k) +  u F^2 (k - P) \right] .
\label{phi_pion_integral}
\ee
The expression for the normalization constant, the weak pion decay
constant, $F_\pi^2$, has been obtained in Ref.\cite{DP86}.  Note the
additional form factors, $F(k) F(k - P)$, in the integrand in
Eq.(\ref{phi_pion_integral}), as compared to the photon wave function,
Eq.(\ref{phi_photon_integral}). These are the form factors originating
from the pion--quark coupling, Eq.(\ref{quark_pion_coupling}).
\par
The evaluation of the integrals defining the photon and pion wave
functions, Eqs.(\ref{phi_photon_integral}) and (\ref{phi_pion_integral}),
proceeds as follows. First the integral over $k^+$ is taken, using up
the delta function. Then the integral over $k^-$ is performed by contour
integration. A special property of the light--like coordinates is that
the denominators are linear in $k^-$. As shown in Ref.\cite{PP97}, the
condition that the poles lie on different sides of the real axis ensures
that Eqs.(\ref{phi_photon_integral}) and (\ref{phi_pion_integral}) are
non-zero only for $0 < u < 1$. In the last step the integral over
transverse momenta is computed, taking into account the form factors.
\par
{\em End--point behavior of the wave functions.} In the expressions for 
the photon and pion wave
functions in the effective theory, Eqs.(\ref{phi_photon_integral}) and
(\ref{phi_pion_integral}), the integral over transverse momenta contains
a logarithmic divergence which is regularized by the form
factors. Let us analyze the integrands in order to see, for a given value
of $u$, which regions of $k^\perp$ make the main contribution to the 
integrals. For this we consider the virtualities of the quark
propagators in the loop integrals in the vicinity of the two poles in
$k^-$. Taking into account that $k^+ = u P^+$ one easily sees that
at the pole in $k^-$ corresponding to $k^2 = M^2$ the virtuality
of the quark with momentum $k - P$ is
\be
(k - P)^2 &=& -\frac{|k^\perp |^2 + M^2}{u} ,
\label{virtuality_1}
\ee
while at the pole in $k^-$ corresponding to $(k - P)^2 = M^2$ the 
virtuality of the other quark is
\be
k^2 &=& -\frac{|k^\perp |^2 + M^2}{1 - u} .
\label{virtuality_2}
\ee
We see that the kinematical boundaries, $u \rightarrow 0$ and
$u \rightarrow 1$, correspond to the situation that one of the quarks
has a large space--like momentum.  In these limits the form factors in
the integrand play a crucial role, since they suppress the contributions
of large space--like momenta. More precisely, for values of $u$
parametrically of the order
\be
u &\sim& (M\bar\rho )^2 \hspace{1cm} \mbox{or} \hspace{1cm}
1 - u \;\; \sim \;\; (M\bar\rho )^2
\label{parametric_range}
\ee
one of the quarks has a virtuality of the order $\bar\rho^{-2}$ and the
integral is cut by the form factors already at transverse momenta
of order $|k_\perp | \sim M$. For values of $u$
not close to the boundaries the integral over transverse
momenta extends up to $\bar\rho^{-1}$, leading to the usual
logarithmic dependence of the integral on the ultraviolet cutoff, 
$\bar\rho^{-1}$.
\par
In the light of the above it is clear that the photon and the pion wave
functions, Eqs.(\ref{phi_photon_integral}) and
(\ref{phi_pion_integral}), behave differently at the end points. In the
case of the pion, due to the form factors $F(k) F(k - P)$ introduced by
the coupling of the pion field to the quarks,
Eq.(\ref{quark_pion_coupling}), the contribution from large virtualities
are suppressed, leading to the vanishing of the wave function at the end
points $u \rightarrow 0$ and $1$. In the case of the photon, on the
other hand, the multiplicative factors are absent, and the integral is
not suppressed for $u \rightarrow 0$ and $1$. The remaining form
factors in Eq.(\ref{phi_photon_integral}), which originate from the
momentum dependence of the dynamical mass in the quark propagators, do
not suffice to make the integral go to zero at $u \rightarrow 0$ and
$1$.  Hence there is no reason for the photon wave function to go to
zero at the boundaries.
\par
{\em Numerical estimates.}
To perform a numerical estimate of the pion and photon wave functions we
need to put in the specific form of the form factor, $F(k)$.  This 
function has been derived for Euclidean ({\em i.e.}, space--like) momenta 
as the Fourier transform of the instanton zero mode \cite{DP86}.  One
possibility would be to compute moments of the wave functions, which can
be expressed as integrals over Euclidean momenta.  However, we would
like to compute the wave function directly, since, for instance, very high
moments would be needed in order to restore the end--point behavior of
the wave function. Thus we prefer to carry out the integrals
Eqs.(\ref{phi_photon_integral}) and (\ref{phi_pion_integral}) over
Minkowskian momenta. In principle one could continue the exact Fourier
transform of the instanton zero mode to Minkowskian momenta; the
function exhibits a cut at positive Minkowskian $k^2 > 0$. Since the
numerical evaluation of the integrals with this function is rather
tedious, we shall instead use a simple pole form,
\be
F(k) &\rightarrow& \frac{\Lambda^2}{\Lambda^2 - k^2 - i0}
\label{F_pole}
\ee
($k^2$ is the Minkowskian momentum). With
$\Lambda^2 \sim 2.0 \, \bar\rho^{-2}$ this form gives a good overall
approximation to the Fourier transform of the zero mode in the Euclidean
domain ($k^2 < 0$). With the form factors approximated by
Eq.(\ref{F_pole}), the integrals Eqs.(\ref{phi_photon_integral}) and
(\ref{phi_pion_integral}) can be evaluated by contour
integration over $k^-$. We emphasize that the prescription for dealing with the
poles of the form factors, {\em cf.}\ Eq.(\ref{F_pole}), follows unambiguously
from the requirement that the moments of the wave function computed with
Eq.(\ref{F_pole}) coincide with the corresponding Euclidean
integrals. Since we have seen above that the endpoint behavior of the
wave function is governed by the form factor at large space--like
momenta, {\em cf.}\ Eqs.(\ref{virtuality_1}) and (\ref{virtuality_2}), 
where the pole form Eq.(\ref{F_pole}) is a good approximation to the exact
Fourier transform, we are confident that the use of Eq.(\ref{F_pole}) is
at least qualitatively correct.
\par
Dorokhov has discussed the pion electromagnetic form factor 
in connection with the instanton vacuum using a dispersion relation 
approach \cite{Dorokhov96}. He quotes an expression for the pion
wave function which involves the zero mode form factor 
corresponding to the instanton in regular gauge, which is not consistent 
with the superposition of instantons (sum ansatz) implied in the derivation
of the instanton medium \cite{DP86}. Furthermore, in the language of
our approach, his result apparently amounts to neglecting the contributions 
from the singularities of the form factors $F(k)$ in the Minkowskian loop 
integral, and thus seems to have no clear relation to a Euclidean calculation 
of moments of the wave function.
\par
\begin{figure}
\setlength{\epsfxsize}{15cm}
\setlength{\epsfysize}{15cm}
\epsffile{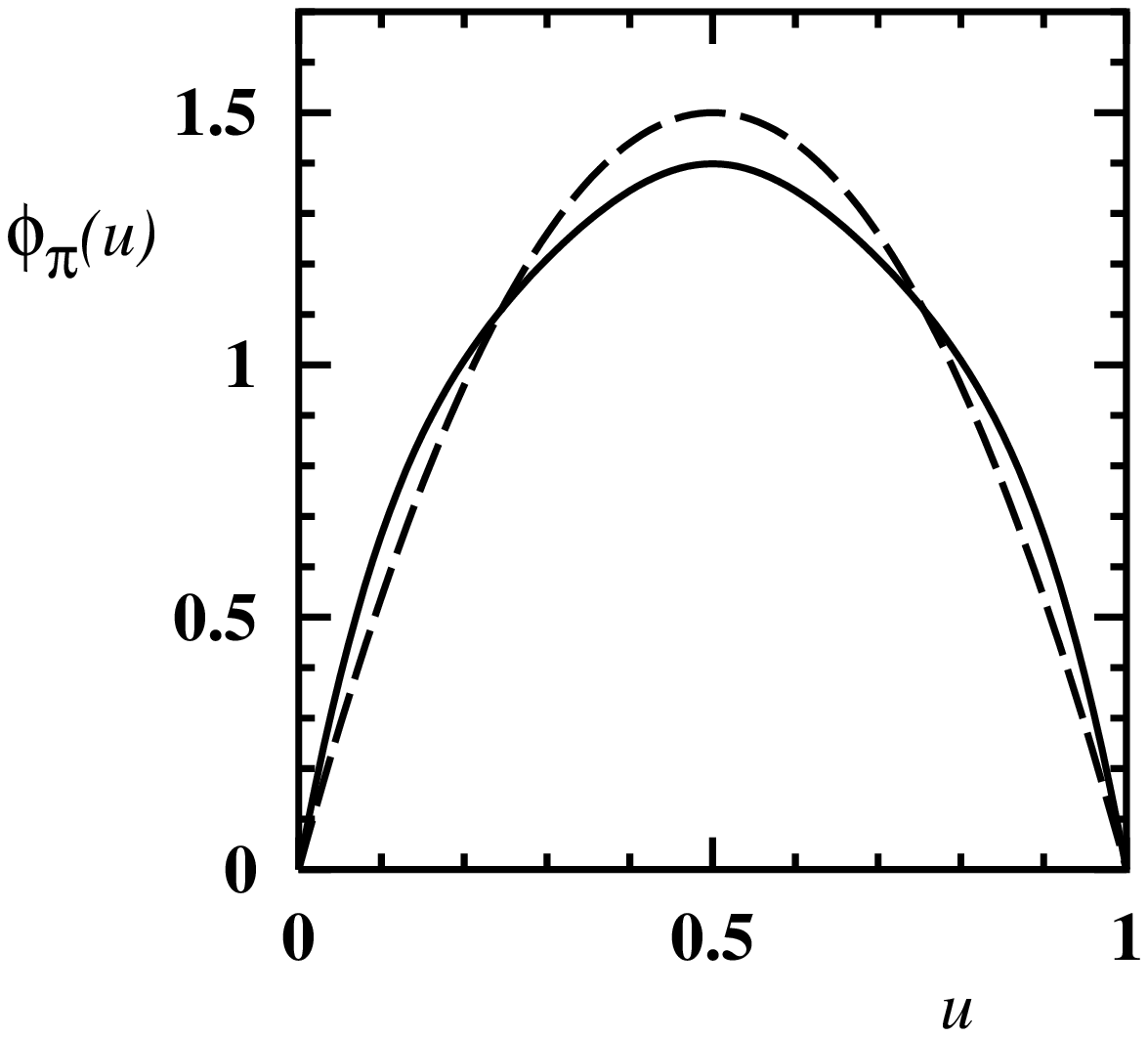}
\caption[]
{The pion wave function, $\phi_\pi (u)$ {\it Solid line:}
Wave function calculated in the low--energy effective theory,
{\it cf.}\ Eq.(\ref{phi_pion_integral}). {\it Dashed line:}
Asymptotic wave function, 
$\phi_\pi^{\rm asymp.} (u) = 6 u (1 - u)$.}
\label{fig_fig1}
\end{figure}
\begin{figure}
\setlength{\epsfxsize}{15cm}
\setlength{\epsfysize}{15cm}
\epsffile{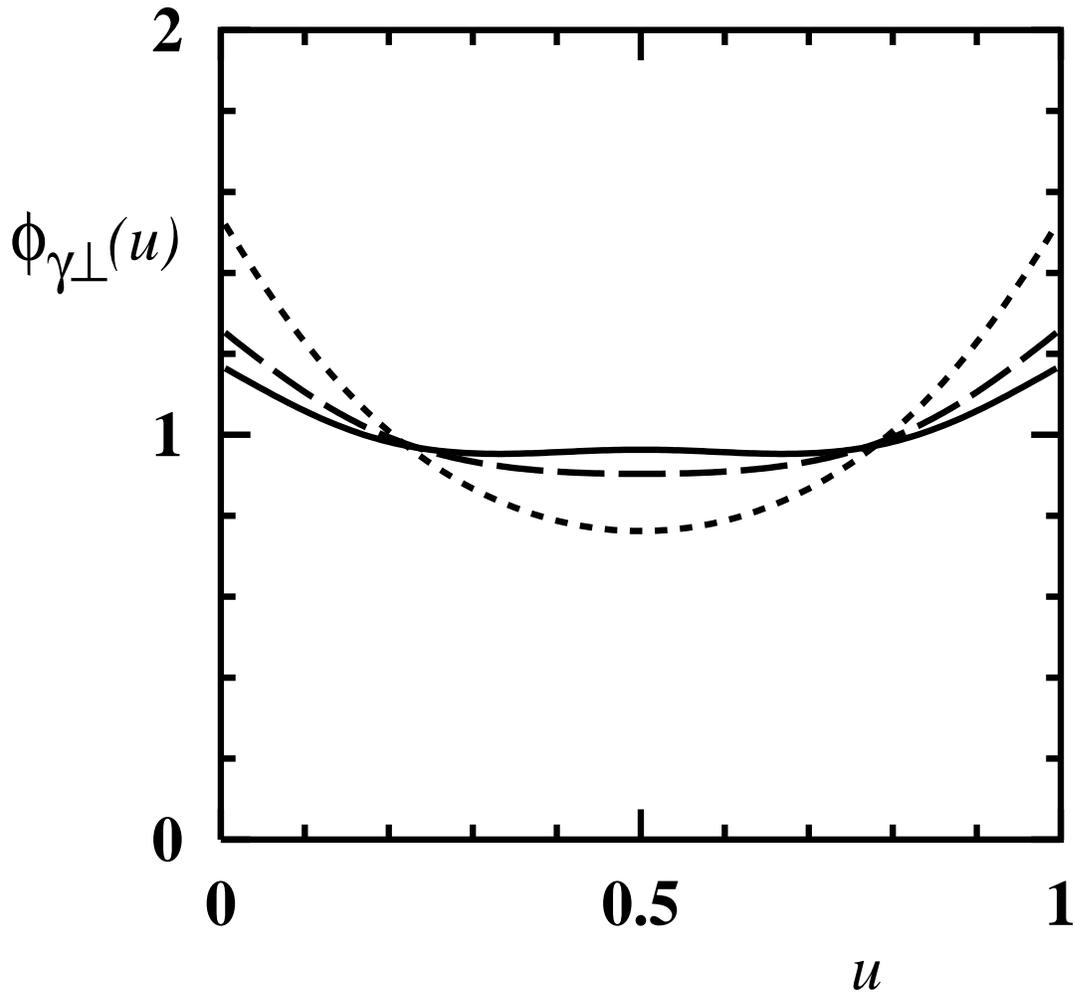}
\caption[]
{The photon wave function, $\phi_{\gamma\perp} (u)$, 
calculated in the low--energy effective theory,
{\it cf.}\ Eq.(\ref{phi_photon_integral}). {\it Solid line:}
real photon ($P^2 = 0$); {\it dashed line:} the corresponding function
for a spacelike virtual photon with $P^2  = -(250\,{\rm MeV})^2$;
{\it dotted line:} $P^2  = -(500\,{\rm MeV})^2$.}
\label{fig_fig2}
\end{figure}

For the numerical estimates we use the standard parameters of the
instanton vacuum ($M = 350\, {\rm MeV}, \, \bar\rho = 600\,{\rm MeV}$).
The results for the pion and photon wave function are shown in
Figs.\ref{fig_fig1} and \ref{fig_fig2}. As can be seen from
Fig.\ref{fig_fig1}, the pion wave function obtained from
Eq.(\ref{phi_pion_integral}) vanishes at the end points, in agreement
with the general argument presented above.
The wave function at the low normalization point is only slightly flatter 
than the asymptotic one\footnote{In
the calculation of the pion wave function in Ref.\cite{PP97} the form
factors inside the square bracket in Eq.(\ref{phi_pion_integral}) and
in the denominators of the quark propagators were
put to unity, since they are not essential for cutting off the
integral. Here we take into account also those factors in
Eq.(\ref{phi_pion_integral}). Our numerical results are
nevertheless very close to those of
Ref.\cite{PP97}.} \cite{BL79,EfrRad80,ChZh77},
\be
\phi_\pi^{\rm asymp} (u) &=& 6 u (1 - u) ,
\label{asymp}
\ee
and far from the form suggested by Chernyak and Zhitnitsky,
$\phi_\pi^{\rm CZ} (u) \, = \, 30 u (1 - u) (2 u - 1)^2$ \cite{ChZh84}.
\par
In order to determine the scale dependence of the wave function one needs to
expand it in eigenfunctions of the evolution equations. To one loop
accuracy, both for the pion \cite{EfrRad80} and the photon \cite{ShifVysot81}
wave functions, Eqs.(\ref{phi_pion}) and (\ref{phi_photon_trans}),
the eigenfunctions are Gegenbauer polynomials of index $3/2$.
The expansion coefficients are given by the moments
\be
B_n &=& N_n^{-1} \int_0^1 du \; C_n^{3/2} (2u -1 ) \phi_\pi (u ) ,
\nonumber \\
N_n &=& \int_0^1 du \; [C_n^{3/2} (2u -1 )]^2 \phi_\pi^{\rm asymp} (u) .
\label{B_n}
\ee
Computing these coefficients for the pion wave function obtained from
the effective low--energy theory, Eq.(\ref{phi_pion_integral}), we find
the expansion
\be
\phi_\pi (u) &=& \phi_\pi^{\rm asymp} (u)
\left[ 1 \; + \; 0.062 \, C^{3/2}_2 (2 u - 1) \; + \;
0.01 \, C^{3/2}_4 (2 u - 1)
\; + \; \ldots \right].
\label{pion_gegenbauer}
\ee
Our value for the second moment, $B_2 = 0.062$, is considerably smaller
than that of Chernyak and Zhitnitsky, $B_2 = 0.66$.  One notes that the
coefficient of the fourth--order polynomial is already very small
numerically.  
We remark that Eq.(\ref{pion_gegenbauer}) provides also a
reasonable numerical representation of the computed wave function.
\par
The recent CLEO measurements \cite{CLEO98} of the transition form factor
$\gamma\gamma^\ast \rightarrow \pi^0$ provide a unique opportunity
to extract information about the pion wave function. In the standard
hard scattering approach ({\em i.e.}, using the tree--level coefficient
function) the $1/Q^2$--asymptotic behavior of the transition form factor
is goverened by the ``inverse'' moment of the pion wave 
function \cite{BL79,JKR96},
\be
I(\mu^2 ) &=& \int_0^1 du \, u^{-1} \, \phi_\pi (u,\mu^2) ,
\label{inverse_moment}
\ee
where $\mu\simeq Q$. With the asymptotic wave function,
Eq.(\ref{asymp}), one obtains a value of $I^{\rm asymp} = 3$, while
the Chernyak--Zhitnitsky wave function at the initial normalization
point gives $I^{\rm CZ} = 5$ \cite{ChZh84}. With the wave function
computed in the effective low--energy theory, Eq.(\ref{phi_pion_integral}),
we find a value of $I = 3.21$, which should be associated with 
a normalization point of the order of 
$\mu \simeq \bar\rho^{-1} = 600\, {\rm MeV}$.
In Table \ref{table_1} we give the values of the integral 
Eq.(\ref{inverse_moment}) obtained by leading--order
evolution ($\Lambda_{\rm QCD} = 250\,{\rm MeV}$, $N_f = 3$) 
of our wave function, {\em cf.}\ Eq.(\ref{pion_gegenbauer}),
at a few values of experimentally relevant scales. 
[For details concerning the evolution see Ref.\cite{ChZh84}.]
We note that the values are close to those obtained in a QCD sum 
rule approach with non-local condensates \cite{BakMikh98}.
\par
It is known that the inclusion of $\alpha_s$ corrections to the
coefficient function decreases the coefficient of the
$1/Q^2$--asymptotic behavior of the transition form factor by about 
15--20\%, see Refs.\cite{KrollRaulfs96,MusRad97}.
Including these corrections our value for $I$
is consistent with the CLEO results, while that of Chernyak and 
Zhitnitsky seems to be ruled out \cite{CLEO98}.
Note also that our value is comparable with the one
extracted from a QCD sum rule for the
form factor $\gamma\gamma^\ast \rightarrow \pi^0$ \cite{RR96}.
\begin{table}
\begin{center}
\begin{tabular}{r|cccccc}
$\mu^2 / {\rm GeV}^2 $ &
0.6  & 1    & 4    & 8    & 10   & 100 \\ \hline
$I(\mu^2 )$ &
3.21 & 3.16 & 3.12 & 3.11 & 3.08 & 3.06
\end{tabular}
\end{center}
\caption[]{The leading--order scale dependence of the ``inverse moment'', 
Eq.(\ref{inverse_moment}),
obtained with the pion wave function from the effective low--energy
theory (see Fig.\ref{fig_fig1}).}
\label{table_1}
\end{table}
\par
Finally, it is interesting to note that at $u = 1/2$, where the quark and 
antiquark in the pion carry equal momentum fraction,
we obtain a value of the wave function of
\be
\phi_\pi (1/2) &=& 1.4 ,
\ee
which is in good agreement with the bound obtained by Braun and
Filyanov from QCD sum rules in exclusive kinematics,
$\phi_\pi (1/2) = 1.2 \pm 0.3$ \cite{BF89}.
\par
The photon wave function calculated in the effective low--energy theory,
{\em cf.}\ Eq.(\ref{phi_photon_integral}), is shown in
Fig.\ref{fig_fig2}.  It does not go to zero at the boundaries. Thus, the
numerical results support the above general conclusions of different
behavior of the photon and pion wave functions. 
We do not write a representation analogous to Eq.(\ref{pion_gegenbauer})
for the photon wave function. Such an expansion would be meaningless ---
since the function does not vanish at the end points, the moments do not
decrease rapidly, and a very large number of terms would be needed to
represent the function even for values of $u$ not close to the
boundaries.
\par
For the normalization constant of the photon wave function, 
Eq.(\ref{normalization_integral}), we obtain a value of
\be
f_{\gamma\perp} &=& 0.036\,  N_c M \;\; \simeq \;\; 38\,{\rm MeV}
\label{f_gamma}
\ee
at the low normalization point. 
Due to the non-conservation of the tensor current $f_{\gamma\perp}$ is 
actually scale--dependent \cite{ShifVysot81}.
This quantity is directly related to the so--called magnetic susceptibility 
of the quark condensate, $\chi_q$, introduced in 
Refs.\cite{BalYung83,IoffeSmilga84}, namely
$f_{\gamma\perp} = \langle \bar{u} u \rangle \chi_u$. 
One should compare our result with the value obtained in 
Refs.\cite{BKYu85,BeKo84}
from a QCD sum rule approach , $f_{\gamma\perp} = 68\, {\rm MeV}$
at $\mu = 1\,{\rm GeV}$ (using a value of
$\langle \bar{u} u \rangle = -(250\, {\rm MeV})^3$ at $\mu = 1\,{\rm GeV}$).
Assuming a normalization point of 
$\mu \simeq \bar\rho^{-1} = 600\, {\rm MeV}$ for $f_{\gamma\perp}$
calculated in the effective theory the 
value given in Eq.(\ref{f_gamma}) should be reduced by 
a few percent at $\mu = 1\, {\rm GeV}$.
\par
{\em Off-shell behavior of the photon wave function.}
The photon wave function, Eq.(\ref{phi_photon_trans}), is defined as the
matrix element of a twist--2 operator between a physical photon state
($P^2 = 0$) and the vacuum. It is interesting to consider the
corresponding correlation function of the light--cone operators with the
electromagnetic current also at space-like momentum transfers
($P^2 < 0$). We define:
\be
\lefteqn{
\int d^4 x \; e^{-i P\cdot x} \langle 0 | \mbox{T}
\left\{ J_\rho^{\rm e.m.} (x) , \bar u (z)
\sigma_{\mu\nu} [z, -z] u(-z) \right\} | 0 \rangle
} && \nonumber \\
&=& i \, {\textstyle\frac{2}{3}} f_{\gamma\perp} (P^2 )
\left ( g_{\rho\mu} p_\nu - g_{\rho\nu} p_\mu \right)
\int_0^1 du \; e^{i (2 u - 1) p\cdot z} \phi_{\gamma\perp} (P^2, u)
\;\; + \;\; \ldots ,
\label{correlator_trans}
\ee
where $p$ is a light--like vector defined in such a way that it
coincides with $P$ in the limit $P^2 \rightarrow 0$,
\be
p_\mu &=& P_\mu - \frac{P^2}{2 (z\cdot P )} z_\mu ,
\hspace{2cm} p^2 \;\; = \;\; 0 ,
\ee
and we have not written out terms with other tensor structures which
correspond to higher twists. The function $\phi_{\gamma\perp} (P^2, u)$,
which we define to be normalized according to Eq.(\ref{normalization})
also for $P^2 < 0$, reduces to the photon wave function,
$\phi_{\gamma\perp} (u)$ in the limit $P^2 \rightarrow 0$.  In the
effective low--energy theory it is given by the expression
Eq.(\ref{phi_photon_integral}) with $P^2 < 0$.  The numerical results
for momenta $P^2 = -(250 \, {\rm MeV})^2$ and 
$P^2 = -(500\, {\rm MeV})^2$ are shown in
Fig.\ref{fig_fig2}. One sees that the wave function becomes larger at
the boundaries for increasing space--like photon momentum. The result
for the normalization constant is for momenta
$0 < -P^2 < 1 \, {\rm GeV}^2$ well approximated by the form
\be
f_{\gamma\perp} (P^2 ) &\simeq&
f_{\gamma\perp} \left( 1 - \frac{P^2}{2.2\,\bar\rho^{-2}} \right)^{-1} ,
\ee
where $f_{\gamma\perp}$ is given by Eq.(\ref{f_gamma}).
\par
{\em Relation of meson wave functions to off-forward 
parton distributions.}
The description of deeply virtual Compton scattering or hard meson
production requires the so--called off-forward parton distributions
(OFPD's) of the nucleon \cite{R96,Ji97}. A novel feature of the these
compared to the usual parton distribution functions is the dependence on
the longitudinal component of the momentum transfer, $\xi$ (see
\cite{Ji97} for definitions). A convenient language to understand
general aspects of the $\xi$--dependence of the OFPD's, according to
Radyushkin \cite{Radyushkin98}, are the so-called double distributions.
In particular, he discusses a ``meson exchange'' type contribution to
the double distributions which relates to the OFPD's in the kinematical
region $-\xi / 2 < x < \xi / 2$. This argument relates the behavior
of the meson wave function at the boundaries, $u = 0, 1$, to that of the
OFPD at $x = \pm \xi / 2$. In particular, if the meson wave function at
$u = 0, 1$ were non-zero, the OFPD would be discontinuous at
$x = \pm \xi / 2$, which would spoil the factorization of the amplitude.
\par
Strong variations of the OFPD of the nucleon near $x = \pm \xi / 2$ have
been observed in a calculation in the effective low--energy theory in
the large--$N_c$ limit, where the nucleon is described as a chiral
soliton \cite{PPPBGW97}. It was seen there that near $x = \pm \xi / 2$
the behavior of the OFPD is governed by the momentum dependence of the
dynamical quark mass, which turns a would--be discontinuity into a sharp
but continuous crossover. This is consistent with the observation made
in the above calculation of wave functions, namely that it is the
momentum dependence of the dynamical quark mass that determines also the
end--point behavior of the meson wave function. One important difference
between the wave functions and the OFPD's in this approach is due to the
role of the formal parameter $N_c$ (number of colors): While in the case
of the meson wave function the parametric range of those values of $u$
close to the boundaries essentially affected by the momentum--dependent
dynamical quark mass is given by Eq.(\ref{parametric_range}), in the
case of the OFPD the crossover region in $x$ near $\pm\xi / 2$ where the
momentum--dependent mass is essential is parametrically of the order
\cite{PPPBGW97}
\be
|x|-\xi/2 &\sim&  \frac{(M\bar\rho)^2 M}{M_N}
\;\; \sim \;\; \frac{(M\bar\rho)^2}{N_c} .
\label{crossover_region}
\ee
The ``crossover'' region of the OFPD is parametrically smaller than the
``boundary'' region of the wave function.  Nevertheless, the physical
mechanism --- the suppression of large quark virtualities due to the
momentum--dependent quark mass --- is the same in both cases. Thus the
effective theory derived from the instanton vacuum, with the ensuing
fully field--theoretic description of the nucleon as a chiral soliton,
provide a consistent realization of the general relations noted in
Ref.\cite{Radyushkin98}.
\par
{\em Conclusions and outlook.}
In this paper we have computed the photon and pion wave function in the
effective low--energy theory derived from the instanton vacuum. We have
exhibited the reason for the vanishing of the pion wave function at the
end points --- the suppression of large quark virtualities by the
momentum--dependent dynamical quark mass --- and seen that the
corresponding mechanism is absent in the 
photon\footnote{Qualitative arguments in favor of an important role of
the momentum--dependent quark mass in hadron wave functions have 
been presented in \cite{Zhit97}.}. 
\par
As to the numerical reliability of the calculated wave functions, we
would like to take a very modest point of view. There is an intrinsic
uncertainty in the parameters of the effective low--energy theory,
related to the approximations made in the instanton model of the QCD
vacuum, which is based on the smallness of the packing fraction,
$\bar\rho / \bar R \simeq 1/3$.  Nevertheless, our qualitative
conclusions concerning the different behavior of the photon and pion
wave functions stand up, since they follow from the general structure of
the dynamical quark mass and the quark--pion coupling in the effective
low--energy theory, which is unambiguous at least to leading order in
$\bar\rho / \bar R$.
\par
Our result for the pion wave function at the low normalization point
is close to the asymptotic form and consistent with the CLEO measurements.
The fact that we obtain a shape substantially different from
the Chernyak--Zhitnitsky one is due to a significantly smaller value of the 
second moment, and, more importantly, the taking into account of {\em all}
moments of the wave function (which is to say, the avoidance of working
with explicit moments) by our approach. In this sense our results
support conclusions reached previously in Refs.\cite{MR86,BF89}.
\par
We have pointed out that the physical mechanism determining the
end-point behavior of the meson wave function and the behavior of the
off-forward parton distribution at the transition points $x = \pm\xi /
2$ are the same --- the momentum dependence of the dynamical quark
mass. The fact that the low--energy effective theory allows to calculate
both quantities in a consistent framework makes it a particularly
valuable tool for investigating the unknown off-forward distributions.
\par
The effective chiral theory, Eq.(\ref{L}), allows to compute 
also the light--cone wave functions of many--pion states.
Recently the two--pion wave function has been studied in this approach,
which is needed to describe exclusive pion production in processes such 
as $\gamma^\ast\gamma \rightarrow \pi\pi$ or
$\gamma^\ast p \rightarrow p + 2\pi , 3\pi$ {\em etc.} 
\cite{PW98}.
\par
The approach outlined in this paper can be extended to study also the
higher--twist components of the meson and photon wave functions. In this
case, however, one has to take into account also explicit contributions
from the path--ordered exponentials of the gauge field,
Eq.(\ref{P_exp}). This can be done using the method of effective gluon
operators in the instanton vacuum developed in Refs.\cite{DPW96}.
\\[.5cm]
{\large\bf Acknowledgements} \\[.2cm]
The authors are grateful to P.V.\ Pobylitsa for his help during the
initial stages of this work, and to N.G.\ Stefanis for discussions 
and valuable suggestions.
\\[.2cm]
This work has been supported in part by a
joint grant of the Russian Foundation for Basic Research (RFBR) and the
Deutsche Forschungsgemeinschaft (DFG) 436 RUS 113/181/0 (R), by
RFBR grant 96-15-96764, by the NATO Scientific Exchange grant
OIUR.LG 951035, by INTAS grants 93-0283 EXT and 93-1630-EXT, 
by the DFG and by COSY (J\"ulich).
R.R.\ was supported partially by RFBR Grant 96-02-17631.

\newpage
%
%
\newpage
\newpage
\end{document}